\documentclass[final,a4paper,11pt,3p,twocolumn,times]{elsarticle}

\usepackage{graphicx}
\usepackage{amsmath, amssymb, amsfonts,wasysym}
\usepackage[colorlinks,citecolor=blue]{hyperref}
\usepackage[american]{babel}

\title{On the Evolution of BM Orionis}
\author[itb]{R. Priyatikanto\corref{cor1}}
\ead{rpriyatikanto@students.itb.ac.id}
\cortext[cor1]{Corresponding author}
\address[itb]{Prodi Astronomi Institut Teknologi Bandung, Jl. Ganesha no. 10, Bandung,\\ Jawa Barat, Indonesia 40132}

\journal{Jurnal Matematika dan Sains}

\usepackage{fancyhdr}
\newcommand{\msun}{M$_{\odot}$}

\newcommand{\rsun}{R$_{\odot}$}

\begin{document}
\begin{abstract}
BM Orionis, eclipsing binary system that located in the center of Orion Nebula Cluster posses several enigmatic problems. Its intrinsic nature and nebular environment make it harder to measure the physical parameters of the system, but it is believed as Algol type binary where secondary component is pre-main sequence star with larger radius. To assure this, several stellar models ($M_1=5.9$ {\msun} and $M_2=2.0$ {\msun}) are created and simulated using MESA. Models with rigid rotation of $\omega=10^{-5}$ rad/s exhibit considerable similar properties during pre-main sequence stage, but 2.0 {\msun} at assumed age of $\sim10^6$ is $6.46$ times dimmer than observed secondary star. There must be an external mechanism to fill this luminosity gap. Then, simulated post-main sequence binary evolution of BM Ori that involves mass transfer shows that primary star will reach helium sequence with the mass of $\sim0.8$ {\msun} before second stage mass transfer.
\vskip5pt
\noindent \emph{\textbf{Keywords}: binary star, stellar and binary evolution}
\end{abstract}

\maketitle

\section{Introduction}
Orion (known as \emph{Waluku} in Javanese) is special constellation for people around equator and becomes an icon for equatorial heaven. In the heart of this constellation, there is a young embedded Orion Nebula Cluster (ONC) with its exotic trapezium stars ($\theta^1$ Ori) as an evident of mass segregation in young cluster \cite{hillendbrand97}. Aggregates of relatively more massive stars in the center form hierarchical multiple stellar system as already observed clearly \cite{close03}. Among 5 prominence systems ($\theta^1$ Ori A -- E), $\theta^1$ Ori B is interesting with 5 members and Algol type eclipsing binary as its parent/central, known as BM Orionis.

BM Ori (HD37021) which located at $\alpha_{2000}=05^{\text{h}}35^{\text{m}}16.117^{\text{d}}$ and $\delta_{2000}=-05^{\circ}23'6.86"$ is eclipsing binary with B V star as primary and larger but less massive star as secondary component \cite{popper76,vitrichenko00}. Eventhough study about this system converges toward one conclusion about the primary component, physical parameters of the secondary are still uncertain. Light curve with shallow secondary minima and nearly unresolved spectra keeps BM Ori in mystery. Several models have been proposed to explain the observed phenomena, such as secondary star with flatten disk \cite{hall71} and disk shell \cite{huang75} or spherical shell \cite{vitrichenko96} surrounding the primary.

Based on its light curve, it's obvious that BM Ori is detached system \cite{antonkhina89}, but the interaction between components will influence future evolution of the binary and also the stability of multiple system as a whole. Primary component will soon leave main sequence open the mass transfer channel toward its couple. Stellar evolution in this close binary system will be studied.

This article is devided into 5 sections. The conducted observations toward BM Ori and its environment are reviewed in Section 2. Section 3 explains the orbital and physical properties of both primary and secondary component. Section 4 describes the main tool and the results of the simulation. Section 5 gives the closing remarks.

\begin{figure*}[ht]
\centering
\includegraphics[width=0.95\textwidth]{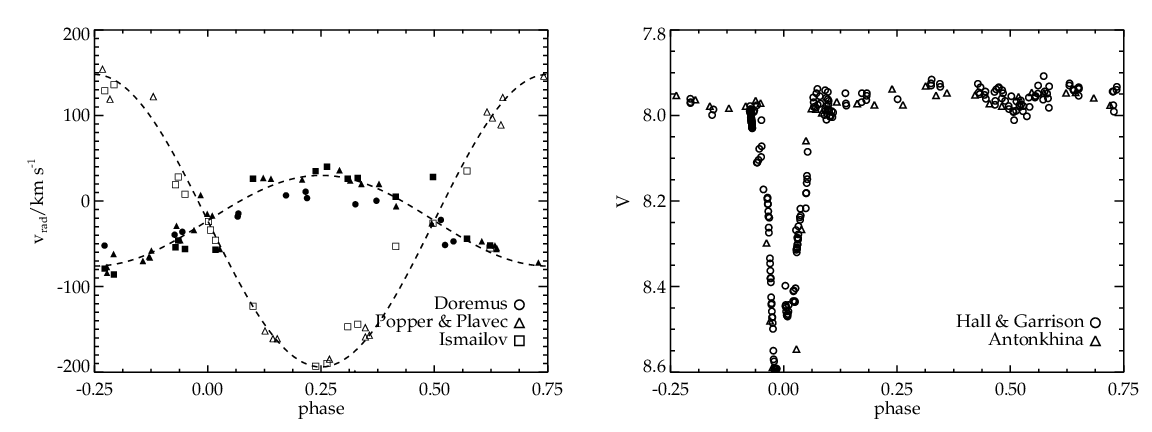}
\caption{Left panel shows radial velocity curve for primary (blue) and secondary component (red) compiled from literatures \cite{doremus70,popper76,ismailov88}. Right panel shows light curve from Hall \& Garrison (1969) and Antonkhina (1989).}
\label{fig:obs}
\end{figure*}

\section{Observations of BM Ori}
Photometric observation of BM Ori and its vicinities have been done intensively since mid-1900s using Johnson UBV filter \cite{hall69,arnold75}. Variability of this eclipsing binary has period of $\sim6.5$ days and amplitude of $\sim0,7$ mag. These early photometric study also concluded that secondary component is 3 times larger that the primary.

Spectroscopic observation of this system has its own challanges which come from intrinsic properties of the binary and the nebular environment of ONC. Although successfully measured radial velocity of primary component, Johnson (1965) and Doremus (1970) haven't identify the spectra of secondary component. Several years later, Popper \& Plavec (1976) used D-line ($\lambda\sim4050\aa$) to measure secondary star's kinematics. Next decade, Ismailov (1988) analysed He I lines that represents primary component and metal lines (Fe I, Ca II and Mg II) that represents the secondary. Emission profile that was found among the metal lines may indicate the presence of a shell.

Spectroscopic observations during eclipse have also been carried out to gain	microturbulence speed and abundances of some important metals \cite{vitrichenko00,vitrichenko01}. On the other hand, the abundace of the primary star just yet determined qualitatively. The only conclusion was that primary star has Helium abundance comparable to the Sun. 

In addition to photometry and spectroscopy, direct imaging and astrometry observations to clarify multiplicity of trapezium stars have been conducted \cite{close03}. From those observations, $\theta^1$ Ori B is confirmed as multiple stellar system with (at least) 5 member, BM Ori ($\theta^1$ Ori B1) becomes the central body, surrounded by B2, B3 and less massive B4.

\section{Properties of BM Ori}
\subsection{Orbital Properties}
BM Ori located in the heart of ONC, 418 pc away from the Sun \cite{menten07}. This eclipsing binary has nearly circular orbit with period of 6.470525 days \cite{hall69} and orbital separation of $\sim30$ {\rsun}. Here is the ephemeris of BM Ori:
\begin{equation}
T_{\text{min}}=JD2440265.343+6.470525-E
\end{equation}

This eclipsing binary with nearly $90^{\circ}$ inclination is believed as detached system which experiences partial eclipse. The \emph{Roche lobe} filling factor of this system is 0.16 and 0.90 for primary and secondary component respectively \cite{antonkhina89}. Table \ref{tab:orbit} summarizes orbital parameters from literatures.

\begin{table*}
\centering
\caption{Orbital parameters of BM Ori from literatures that consist of orbital separation ($a$), eccentricity ($e$), inclination ($i$) and mass ratio ($q$).}
\label{tab:orbit}
{\renewcommand{\tabcolsep}{0.2cm}
\renewcommand{\arraystretch}{1.2}
\begin{tabular}{lccccl}
\hline
Reference 				& $a$(\rsun) 	& $e$ 		& $i$ ($^{\circ}$)& $q$	& metode \\
\hline
Struve \& Titus (1994)	& \ldots		& $0.14$		& \ldots		& \ldots	& spectroscopy\\
Parenago (1957) 		& $50$		& $0.14$		& $87.7$		& $0.25$	& spectroscopy\\
Hall \& Garrison (1969) 	& $32\pm3$	& \ldots		& $83.8\pm2.1$& $0.52$	& phtometry\\
Popper \& Plavec (1976)	& $29.0\pm1.5$& \ldots		& $83\pm4$	& $0.30$	& spectroscopy\\
AlNaimy \& AlSikab (1983)	& $29$		& \ldots		& \ldots		& $0.55$	& phtometry\\
Ismailov (1988)			& $29$		& $0.15$		& \ldots		& $0.37$	& spectroscopy\\
\hline
\end{tabular}}
\end{table*}

\begin{table*}
\centering
\caption{Physical parameters of BM Ori's primary component from literatures, which are Hall \& Garrison (1969) [HG69], Popper \& Plavec (1976) [PP76], Antonkhina et al. (1989) [An89].}
\label{tab:prim}
{\renewcommand{\tabcolsep}{0.2cm}
\renewcommand{\arraystretch}{1.2}
\begin{tabular}{l|ccc|c}
\hline
Parameter		& HG69			& PP76			& An89		& Adopted\\
\hline
MK class		& B2--B3			& B3V			& B2V		& B3V\\
$(B-V)_0$		& $0.08$			& $-0.21\pm0.02$	& \ldots 		& $-0.21$\\
$M_V$		& $0.7$			& $-0.8\pm0.3$		& \ldots 		& $-0.70$\\
$T_{\text{eff}}$& $18700-22000$	& $18700$		& $22000$	& $18700$\\
$R_{\text{eq}}/$\rsun & $2.5$		& $3.0\pm0.4$		& $2.1$		& $3.0$\\
oblateness	& \ldots			& $0.87$			& $1.00$		& $1.0$\\
$M/$\msun	& $5.4$			& $5.9\pm0.8$		& $5.9\pm0.9$	& $5.9$\\
abundance	& \ldots 			& \ldots 			& \ldots 		& $Z=0.02$\\
$v_{\text{rot}} [km/s]$ & \ldots 		& $300$			& \ldots 		& $300$\\
\hline
\end{tabular}}
\end{table*}

\begin{table*}
\centering
\caption{Physical parameters of the secondary compiled from literatures such as Hall \& Garrison (1969) [HG69], Popper \& Plavec (1976) [PP76], Antonkhina et al. (1989) [An89], Vitrichenko \& Plachinda (2000) [VP00], Vitrichenko \& Klochkova (2001) [VK01].}
\label{tab:seku}
{\renewcommand{\tabcolsep}{0.2cm}
\renewcommand{\arraystretch}{1.2}
\begin{tabular}{l|cccc|c}
\hline
Parameter		& HG69		& PP76			& An89		& VP00	 & Adopted\\
\hline
MK class		& A1			& A5--F0			& A3--A4		& G2III	& A5-A6\\
$(B-V)_0$		& $0.07$		& $0.17\pm0.10$	& \ldots 		& \ldots& $0.17$\\
$M_V$		& $-1.1$		& $0.2\pm0.4$		& \ldots 		& \ldots	& $-0.55$\\
$T_{\text{eff}}$& $9400$	& 	$7200-8200$		& $9020$		& 5740	& $8000$\\
$R_{\text{eq}}/$\rsun & $8.5$	& $7.0\pm0.1$		& $8.0$		& $2.5$	& $7.0$\\
oblateness	& \ldots		& $0.57$			& $0.74$		& \ldots & $1.0$\\
$M/$\msun	& $2.8$		& $1.8\pm0.4$		& $2.15\pm0.4$& $2.5$& $2.0$\\
abundance	& \ldots & \ldots & \ldots & $[M/H]=-0.5^{\text{dex}}$ & $Z=0.02$\\
$v_{\text{rot}} [km/s]$ & \ldots 	& $50-100$		& \ldots 		& $60$ & $60$\\
\hline
\end{tabular}}
\end{table*}

\subsection{Primary Component}
Primary component has visual magnitude of $V=8.37$ and intrinsic color of $(B-V)_0=-0.21$, $(U-B)_0=-0.80$ after correction using $E_{B-V}=0.30$. Assuming distance modulus of 8.2, Popper \& Plavec (1976) derived absolute magnitude of $M_V=-0.80$ and confirmed that the primary is B2-3 type main sequence star. More accurate distance ($d=418$ pc) from parallax measurement \cite{menten07} and assuming $A_V=3.0E_{B-V}$, the absolute magnitude of this star can be recalculated.
\begin{equation}
M_V=m_V+5-5\log(d)-A_V
\end{equation}

Then, using $T_{\text{eff}}=18700$ K from spectroscopic observations, it has bolometric correction of $BC=1.94$ \cite{schmidt82}. Implying:
\begin{align*}
M_{\text{bol}}&=M_V-BC=-2.64\\
\log(L_1/L_{\odot})&=2.94\\
R_1/R_{\odot}&=2.85
\end{align*}
This derived value is in agreement with derived values from light curve analysis which range from 2.1{\rsun} \cite{antonkhina89} and 3.0{\rsun} or $3.4\pm0.6$ which is derived from observed survace gravity \cite{popper76}.

As expected, the primary component is a fast rotating star with velocity of $250-300$ km/s \cite{popper76}, but still below its critical velocity ($\sim1000$ km/s). This B star also blows stellar wind, responsibles with observed He I and metal emission lines \cite{ismailov88}. X-ray source that coincides with BM Ori (COUP 778) can be explained by shock wind mechanism related to that stellar wind \cite{stelzer05}. Tabel \ref{tab:prim} summarize physical parameters of the primary.

\subsection{Secondary Component}
Secondary component of BM Ori is so hard to be observed that its physical parameters is not well determined. This star is believed to be pre-main sequence star with $V=8.52$ that experiences gravitational contraction. Popper \& Plavec (1976) got $M_V=0.2\pm0.4$ and $(B-V)_0=0.17_{-0.03}^{+0.10}$, while Hall \& Garrison (1969) got $(U-B)_0=-0.02$ which indicates ultraviolet excess of $\delta_{U-B}=0.7$ as observed in another pre-main sequence star. Radio observation and detection of non-thermal emission from BM Ori may related to flare activity of the secondary \cite{felli93}.

Color index of $(B-V)_0=0.17$ corresponds to main sequence star with temperature of $7200-8200$ K (A5-F0), but Antonkhina et al. (1989) derived $T_{\text{eff}}=9020$ K according the light curve while Vitrichenko \& Plachinda (2000) got $T_{\text{eff}}=5740$ K based on the surface gravity. For secondary, determination of surface temperature and spectral class is not easy since the spectrum is not well-resolved from the primary.

Star's magnitude and luminosity can also be recalculated assuming $T_{\text{eff}}\approx8000$ K (average value from literatures) and $BC=-0.14$. It yieds:
\begin{align*}
M_V&=-0.55\\
M_{\text{bol}}&=-0.68\\
\log(L_2/L_{\odot})&=2.17\\
R_2/R_{\odot}&=6.35
\end{align*}

From the orbit, the secondary component has mass of $\sim2$ {\msun} and radius of $\sim7$ {\rsun}. It rotates with velocity of $50-100$ km/s, much slower compared to its pair \cite{popper76,vitrichenko00}. Tidal attraction from the primary causes higher oblateness of this star.

\begin{figure*}[ht]
\centering
\includegraphics[width=0.6\textwidth]{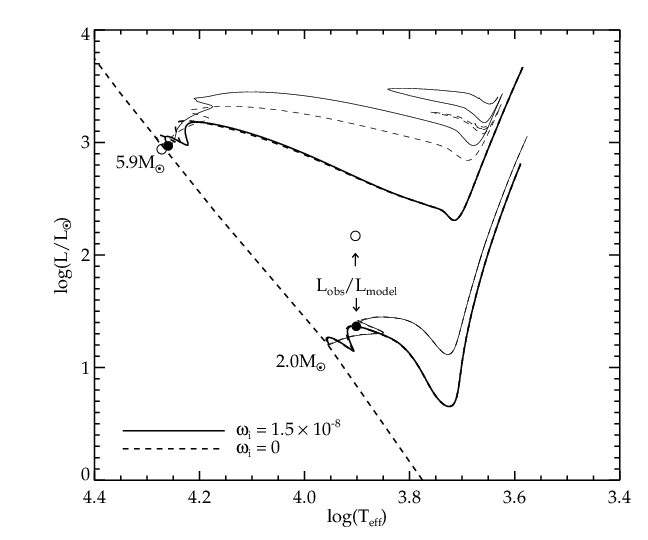}
\caption{Evolutionary track of primary component with mass of $M_1=5.9$ {\msun} (blue) and secondary with mass of $M_2=2.0$ {\msun} (red). Thick and thin lines represent pre and post main sequence stages respectively, while dotted lines belong to rotating models. Filled circles mark the position of assumed model or the best fit toward observed parameters (circles).}
\label{fig:hrdiag}
\end{figure*}

\section{Structure and Evolution of BM Ori}
In this study, the structure and evolution of the stars are simulated using Module for Experiments in Stellar Astrophysics (MESA, \cite{paxton11}). This program is developed according Eggleton's code, adopting updated physical data. This code is constructed using Fortran95 which can be compiled in multi-processor device. Various case of stellar evolution, ranging from pre-main sequence evolution to final collapse of a star can be simulated using this code. Binary evolution that involves mass transfer can also be treated.

Three different models are generated and evolved with MESA. Each model consists of 5.9 {\msun} primary and 2.0 {\msun} secondary component with metalicity of $Z=0.02$ (solar metalicity). The first two models start from Hayashi track with enormous size and luminosity, but with different rotation nature: one without rotation and the other rotates with $\omega_i=1.5\times10^{-8}$ rad/s. This value of angular velocity is choosen in order to make rotating main sequence star with observed rotation velocity ($v_1\approx300$ km/s and $v_2\approx50$ km/s). In these two models, no binary interaction calculated. The last model is the binary evolution model starts from Zero Age Main Sequence (ZMAS) through advanced evolution including mass transfer.

\begin{table*}
\centering
\caption{Global parameter of the models without mass transfer at $\log(t)\approx 6.20$.}
\label{tab:compare}
{\renewcommand{\tabcolsep}{0.2cm}
\renewcommand{\arraystretch}{1.2}
\begin{tabular}{l|cc|cc}
\hline
Parameter	& \multicolumn{2}{c|}{primary}	& \multicolumn{2}{c}{secondary} \\
			& rotating		& non-rotating	& rotating		& non-rotating\\
\hline
$\log(t)$		& $6.200$		& $6.200$		& $6.230$		& $6.209$ \\
$M/$\msun		& $5.900$		& $5.900$		& $2.000$		& $2.000$ \\
$\log(L/L_{\astrosun})$& $2.970$& $3.021$		& $1.366$		& $1.366$ \\
$R/$\rsun		& $3.083$		& $2.978$		& $2.547$		& $2.547$ \\
$T_{\text{eff}}$ [K]		& $18174$	& $19045$	& $7943$		& $7962$ \\
$\log(g)$		& $4.230$		& $4.260$		& $3.927$		& $3.930$ \\
\hline
\end{tabular}}
\end{table*}

\subsection{Evolution Toward Main Sequence}
Departing from Hayashi track, both primary and secondary star contracts to attain new hydrodynamic equilibrium as main sequence star, but with different time scale (less massive star spends more time). During this evolutionary stage, there is no significant difference between non-rotating and rotating models. This result has similar trend compared to rotating model of Martin \& Claret (1996), though their model has smaller mass and faster rotation. Rotating stars have different gravity potential which may influence their internal structure. This is more clearly demonstrated in the post-main sequence evolution.

\subsection{Comparison with Observed Properties}
To make comparison between the model and the observed propertis is not straightforward process since the age of both stars are not precisely determined. Previous study gave a possible range of $10^5-10^6$ years. Primary component has already reached main sequence at age of $8\times10^5$ years. Its structure does not change much during main sequence stage that lasts until the age of $\sim10^7$ years.

The present age of secondary component is harder to approximate because of its pre-main sequence nature. But, as binary component with nearly circular orbit, it is more likely that secondary star formed almost in the same epoch, together with its pair. Although theory of binary star formation doesn't demand simultaneous formation, observation bring evidents toward coevality \cite{brandner97,palla01}. Then age range of $10^6-10^7$ years for both components is reasonable in order to compare model and observation.

As plotted in HR diagram (Figure \ref{fig:hrdiag}), main sequence of 5.9 {\msun} model is rather fit to measured temperature and luminosity of the primary component. Both rotating and non-rotating model show almost similar properties, but rotating model has a bit smaller effective temperature (see Table \ref{tab:compare}).

On the other hand, evolutionary track of 2 {\msun} models are located below the observed properties of secondary star. At the age of $1.6\times10^6$ or $\log(t)=6.2$ both rotating and non-rotating model have the highest luminosity, but still 6.46 dimmer. Difference between these two luminosity demands external processes to occure and add the stellar luminosity. Accumulative reflection from primary component and heating by stellar wind may be sufficiently cover gap \cite{huang75}. Later process is expected to give more contribution.

\begin{figure*}[ht]
\centering
\includegraphics[width=\textwidth]{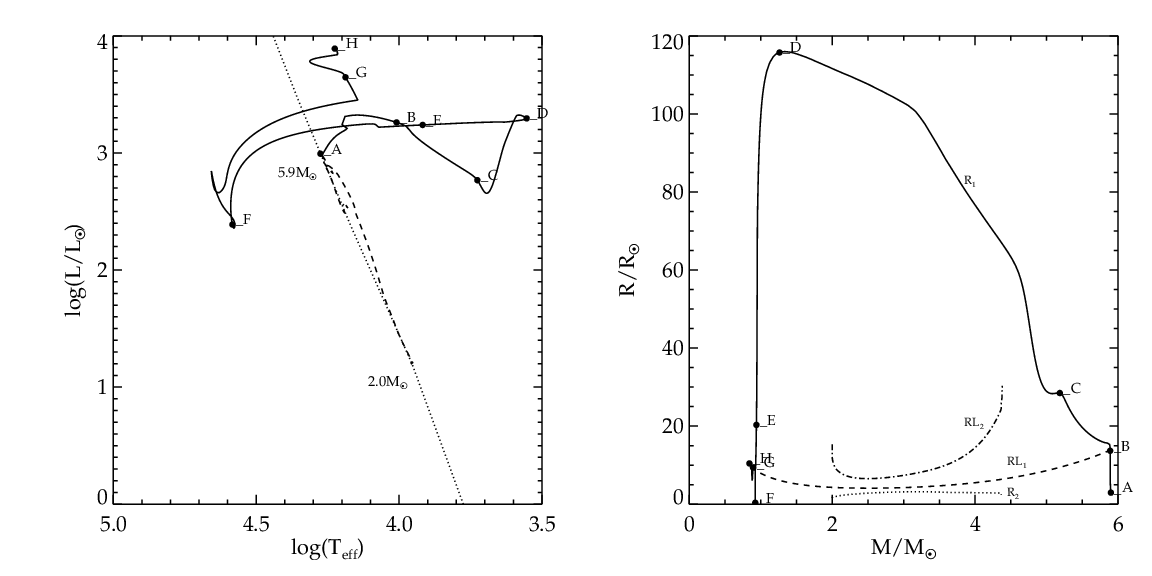}
\caption{\small HR Diagram (left) and mass-radius plot (left) of the donor (blue) and accretor (red). Dashed line in the left panel marks ZAMS while dashed lin in the left marks Roche lobe radius of each star. Letter A--H marks evolutionary stages as described in the text.}
\label{fig:mtrans}
\end{figure*}

\subsection{Post-Main Sequence Evolution}
As relatively close binary system, BM Ori experiences complicated evolution involved mass transfer. After leaving main sequence at age of $\sim10^7$ years, primary component starts to expand, fills its Roche lobe and initiates mass transfer. This post-main sequence mass transfer is an example of case B of Kippenhahn \& Wiegert (1967).

In this model, primary ($M_1=5.9$ {\msun}) and secondary star ($M_2=2.0$ {\msun}) are in the main sequence with similar age. Both stars orbit the center of mass in circular orbit with orbital period of $P=6.47$ days. And here are evolutionary stages experienced by the system:
\begin{enumerate}
\item Primary component leaves main sequence when hydrogen fuel in the center is exhausted. The core shrinks while the envelope expands makes the star fills its Roche lobe (stage B in Fig \ref{fig:mtrans}).
\item Non-conservative mass transfer occurs, sometimes accreting star (secondary) gains same amount of mass transfer from the donor (primary). In this model, mass transfer rate is kept to be constant at $\dot{M}=10^{-5}$ {\msun}/tahun, almost similar to the model of De Greve \& de Loore (1976) for intermediate mass system.
\item Expansion rate of the primary is overwhelmed such that the envelope has much larger radius compared to the orbital separation, common envelope is established. At the same time, star ignites helium burning (stage D).
\item Donor star reaches a new equilibrium as helium star with smaller size and mass ($M_1'=0.8$ {\msun}). On the other hand, accretor becomes more massive ($M_2'=4.4$ {\msun}) while the orbit becomes larger ($a'=40.3$ {\rsun} and $P=13.00$ days).
\item After 10 Gyr, helium star leave its stable condition and expands again. Second stage of mass transfer is initiated (stage G) sets aside smaller mass (stage H) when the simulation is terminated.
\end{enumerate}

\section{Closing Remarks}
In this study, previous observations and studies about BM Ori as an interesting eclipsing binary in the heart of Orion Nebula Cluster are reviewed. However, physical parameters of secondary component are note well-determined. Standard model with assumed parameters  of $M_1=5.9$ {\msun} and $M_2=2.0$ {\msun} does fit with primary component but not for secondary. There must be an external mechanism occurs around the secondary to fill the luminoasity gap.

Simulated binary evolution after main sequence stage shows that mass transfer will transform the donor star to become helium star with stripped envelope. During this stable stage, total mass of the system is around 5.2 {\msun}, much lower that initial total mass. Beside that, orbital parameters are change toward larger separation of $a'=40$ {\rsun} and shorter period of $P'=13$ days. This condition undoubtfully influence the stability of $\theta^1$ Ori B multiple system. Further dynamical analysis need to be done to assess this.

\section*{References}
\bibliographystyle{plainnat}

\end{document}